\documentclass[aps,prb,groupedaddress]{revtex4}
\usepackage{graphicx}

\begin{document}

\newcommand{\be}{\begin{equation}}
\newcommand{\ee}{\end{equation}}
\newcommand{\bea}{\begin{eqnarray}}
\newcommand{\eea}{\end{eqnarray}}
\newcommand{\nt}{\narrowtext}
\newcommand{\wt}{\widetext}

\title{Dissipative phase transition in two-dimensional d-wave Josephson junctions}
\author{D. V. Khveshchenko}
\affiliation{Department of Physics and Astronomy, University of North Carolina,
Chapel Hill, NC 27599}

\begin{abstract}
Quantum dynamics of in-plane 
Josephson junctions between two $d$-wave superconducting films
is described by the anisotropic $XY$-model
where both, quasiparticle and Cooper pair tunneling terms, appear to be equally non-local.
Applying a combination of the weak and strong coupling analyses to this model, 
we find a compelling evidence of a dissipative phase transition. The corresponding 
critical behavior is studied and contrasted with that found previously in
the conventional ($s$-wave) Josephson junctions.
\end{abstract}

\maketitle

Quantum phenomena in ultrasmall Josephson junctions between close 
pairs of superconductors 
have long been at the forefront of research in theoretical and experimental condensed matter physics cite{sz}.
However, putting a few exceptions aside, the previous theoretical 
work was almost exclusively focused on the Josephson junctions 
between conventional $s$-wave superconductors which are fully gapped.
 
The growing interest in unconventional (e.g., $d$-wave) Josephson junctions 
has been bolstered by the studies of cuprates and other gapless superconductors
\cite{otterlo,japan} as well as the proposed applications of 
mesoscopic $d$-wave devices in quantum computing \cite{dwave}.

A unifying theoretical description of tunnel junctions between both, normal metals and superconductors, can be constracted in terms of the 
imaginary-time effective action which governs the dynamics of 
the phase diference across the junction $\phi(\tau)=\phi_R(\tau)-\phi_L(\tau)$ 
conjugate to the junction's charge $q(\tau)$ \cite{sz}.
Under the assumption of weak tunneling, this action can be written in the general form 
\be
S={1\over 4E_c}\int^{1/T}_0({\partial\phi\over \partial\tau})^2d\tau-
\int^{1/T}_0\int^{1/T}_0
[\alpha(\tau_1-\tau_2)\cos{\phi(\tau_1)-\phi(\tau_2)\over 2}
+\beta(\tau_1-\tau_2)\cos{\phi(\tau_1)+\phi(\tau_2)\over 2}]d\tau_1d\tau_2
\ee
where $T$ is the temperature and the first (local) term is the charging 
energy (measured in units of $E_c=e^2/2C$) 
of a junction with capacitance $C$, while the (potentially non-local) $\alpha$- and $\beta$-terms 
represent the processes of quasiparticle and Cooper pair tunneling across the junction, respectively.
The corresponding integral kernels
\bea
\alpha(\tau)=-\sum_{k,k^\prime}|t_{k,k^\prime}|^2G_L(\tau,k)G_R(-\tau,k^\prime)\nonumber\\
\beta(\tau)=\sum_{k,k^\prime}|t_{k,k^\prime}|^2
F_L(\tau,k)F_R(-\tau,k^\prime)
\eea
are determined by the Fourier transforms of the normal and 
anomalous quasiparticle Green functions on the left/right bank of the junction
($G_{L,R}(i\omega_n,k)=(i\omega_n+\xi_{L,R})/(\omega_n^2+\xi^2_{L,R}+\Delta^2_{L,R})$ 
and $F_{L,R}(i\omega_n,k)=\Delta_{L,R}/(\omega_n^2+\xi^2_{L,R}+\Delta^2_{L,R})$, respectively) 
and the tunneling matrix element $t_{k,k^\prime}$.

Under the standard assumption of a generic, 
momentum-non-conserving, tunneling [$t_{k,k^\prime}\approx const$] 
between two fully gapped ($s$-wave) superconductors,
the pair tunneling kernel decays exponentially 
($\beta_s(\tau)\propto\exp(-2\Delta_s\tau)$ for $\tau\gg 1/\Delta_s$). As a consequence,
the constituents of a Cooper pair tunnel almost simultaneously, and the $\beta$-term reduces to
the single time integral $-E_J\int^{1/T}_0\cos\phi(\tau)d\tau$ proportional to
the $\it local$ Josephson energy $E_J=\int^{1/T}_0\beta(\tau)d\tau$. 
 
Although the same argument may seem to be equally applicable to the quasiparticle 
tunneling ($\alpha$-) term, a proper description of the experimental data
on realistic junctions requires one to retain a non-local qusiparticle tunneling term
whose kernel behaves as $\alpha_s(\tau)\propto 1/\tau^2$ 
in the entire range $1/\Delta_s<\tau<1/T$. In this way, one can account for a 
finite subgap resistance due to inelastic (phase-breaking) scattering \cite{sz}.

In contrast, a derivation of the effective action 
(1) of a junction between two gapless $d$-wave (as well as any $l\neq 0$-wave) 
superconductors along the same lines manifests its non-universality
and a strong dependence on the details of the tunneling matrix elements.  
For a momentum-independent tunneling the kernel $\beta(\tau)$ given by Eq.(2) 
vanishes identically, alongside the angular averages
performed in the course of the momentum integrations $\sum_kF_{L,R}(\tau,k)$. 

By contrast, in the presence of a momentum-conserving 
[$|t_{k,k^\prime}|^2\sim\delta^{2}(k-k^\prime)$] 
node-to-node tunneling across the junction between two three-dimensional 
$d$-wave superconductors both kernels in Eq.(1) demonstrate the algebraic decay
$\alpha_{d,3D}\sim\beta_{d,3D}\propto 1/\tau^3$, thereby resulting in the super-Ohmic
quasiparticle dissipative term in Eq.(1) \cite{otterlo,japan}. 
Moreover, the Ohmic quasiparticle tunneling term might still occur for those
relative orientations between the $d$-wave order parameters 
that allow nodal quasiparticles to tunnel directly into the surface-bound 
zero energy states \cite{japan,dwave}.

Interestingly enough, inspite of its being of an equally non-local nature, the
kernel of the concomitant Cooper tunneling term is routinely treated as
though it was local, and it has become customary to endow 
the effective action (1) with the $\it ad$ $\it hoc$ conventional
Josephson energy \cite{otterlo,japan,dwave}.  

The latter approximation may indeed prove
adequate for the junctions formed by 
three-dimensional grains of a $d$-wave superconductor, as, e.g.,
in the case of strongly inhomegeneous high-$T_c$ samples \cite{hm}. 
However, it would obviously fail in the case of tunneling 
between a pair of genuinely two-dimensional 
$d$-wave superconductors (e.g., thin films) where
both the quasiparticle and Cooper kernels show the same Ohmic decay
\cite{joglekar}
\be
\alpha_{d,2D}=\alpha/\tau^2,~~~~~~\beta_{d,2D}=\beta/\tau^2
\ee
In the presence of elastic scattering, this Ohmic behavior of $both$ 
$\alpha(\tau)$ and $\beta(\tau)$ (see Erratum in Ref.\cite{joglekar}) 
will undergo a crossover to the exponential 
decay at time scales in excess of the inverse bulk impurity scattering rate $\gamma$,
thus effectively restoring the local Josephson energy.   
However, as suggested by the wealth of transport data in quasi-planar cuprates,
$\gamma$ turns out to be quite low compared to the maximum gap $\Delta_d$, 
and, therefore, the kernels (3) remain essentially non-local in 
the entire regime $max[1/\Delta_d,1/E_c]<\tau<min[1/T,1/\gamma]$.

Apart from its anticipated relevance to the problem of tunneling between thin 
$d$-wave films, the anisotropic $XY$-model described by 
the effective action (1,3) is interesting in its own right, for
it appears to reside outside the realm discussed in the literature up to date and its 
quantum dynamics has not yet been studied.
To fill in the gap, in what follows we investigate this new model 
by applying a number of complementary techniques
that cover both weak and strong dissipative coupling regimes. 

In the weak coupling regime ($\alpha,\beta<1$), an adequate approach to the model (1,3)
is provided by a direct perturbative expansion
for the grand partition function in the charge representation
\bea
{\cal Z}(Q)=\sum_{n=-\infty}^{\infty}\int^{Q+ne}_{Q+ne}Dq(\tau)
\sum_{N=1}^\infty{1\over N{!}}
\prod_{i=1}^N\int^{1/T}_0d\tau^{+}_i\int^{1/T}_0d\tau^{-}_i
\alpha(\tau^{+}_i-\tau^{-}_i)\nonumber\\
\sum_{M=1}^\infty
{1\over M{!}}\sum_{q_j=\pm 1,\sum q_j=0} 
\prod_{j=1}^M\int^{1/T}_0d\tau^{+}_j\int^{1/T}_0d\tau^{-}_j
\beta(\tau^{+}_j-\tau^{-}_j)\exp(-{1\over 4E_c}\int^{1/T}_0q^2(\tau)d\tau)
\eea
where the lower limit in all the integrals is set at $\tau_c\sim 1/E_c$, 
and the instantaneous value of the total charge of the junction 
\be
q(\tau)=Q+ne+e\sum^N_{i=1}(\theta(\tau-\tau^+_i)-\theta(\tau-\tau_i^-))
+e\sum^M_{j=1}q_j[\theta(\tau-\tau^+_j)+\theta(\tau-\tau_j^-)]
\ee
includes a continuously varying contribution $Q=CV$ induced by an applied external bias $V$.

The sum in (4) accounts for all the trajectories in the charge 
space ($Q+ne\to Q+(n\pm 1)e\to\dots\to Q+ne$) that consist of $N$ 
pairs of quasiparticle ($Q\to Q\pm e$ followed by
$Q^\prime\to Q^\prime\mp e$) and $M$ pairs of Cooper pair
($Q\to Q\pm e$ followed by $Q^\prime\to Q^\prime\pm e$) tunneling events.

The periodic dependence of the partition function (4)
(hence, any physical observable) upon the external charge $Q$ with the period $e$ allows one 
to restrict its values to the "Brillouin zone" (BZ) $-e/2\leq Q\leq e/2$ 
in the charge space. Unlike in the case of the local Josephson energy, 
there is no room for a periodicity with the minimum charge $2e$ even for $\alpha\to 0$.

In the absence of tunneling, the ground and first excited states appear to be degenerate 
($Q^2/2C=(Q\pm e)^2/2C$) at the BZ boundaries ($Q=\pm e/2$).
In the vicinity of the degeneracy points the renormalized gap $\Delta(Q)=E_1(Q)-E_0(Q)$
between the ground and first excited states remains small compared to $E_c$,
and, therefore, one can neglect any transitions 
between these two energy levels and the rest of the spectrum
separated by an energy gap of order $E_c$. As a result, the sum (4) reduces to that
over the trajectories comprised
of a sequence of $\Delta Q=\pm e$ "blips" between the two lowest states \cite{weak}. 

A straightforward analysis shows that this effective two-state 
system can be amenable to the renormalization group (RG)
analysis, akin to that developed in the context of the Kondo and other
quantum spin-$1/2$ problems. In particular, the standard procedure of changing the cut-off in 
the time integrations in (4) from $\tau_c$ to $\tau^\prime_c>\tau_c$
and integrating over all the pairs of opposite blips with separations
$\tau_c<|\tau^{+}_i-\tau^{-}_j|<\tau^\prime_c$ results in the renormalized 
partition function whose relevant part retains the original form (4), albeit with 
renormalized parameter values.

To first order in the dissipative couplings, the wave function renormalization 
$Z$ (defined through the propagators of the two lowest states,
$G_{0,1}=Ze^{-E_{0,1}\tau}$) and the vertex renormalization factor $Z_{\Gamma}$
receive contributions of order $\alpha\ln\tau^\prime_c/\tau_c$ 
and $\beta\ln\tau^\prime_c/\tau_c$, respectively
(notably, the closed Dyson-type equations for the propagators $G_{0,1}(\tau)$
and the vertex function $\Gamma$ contain no analog of the polarization function \cite{weak}).

Combining these results, one arrives at the RG equations
for the effective couplings $\alpha_r=\alpha Z^2Z_\Gamma^2$,
$\beta_r=\beta Z^2Z_\Gamma^2$, and ${\tilde\Delta}_r=\Delta_r\tau_c$
\bea
{d\alpha_r\over d\ln\tau_c}=-2\alpha_r(\alpha_r-\beta_r)\nonumber\\
{d\beta_r\over d\ln\tau_c}=-2\beta_r(\alpha_r-\beta_r)\nonumber\\
{d\ln{\tilde\Delta}_r\over d\ln\tau_c}=(1-2\alpha_r)
\eea
which are valid for as long as the renormalized dissipative couplings
$\alpha_r$ and $\beta_r$ remain small. 

By solving the RG equations (6) and evaluating their solutions
at the lowered energy cutoff $1/\tau^\prime_c=\Delta_r$ 
one obtains the renormalized couplings
\be
\alpha_r={\alpha\beta_r\over \beta}={\alpha\over 1+2\eta\ln E_c/\Delta_r}
\ee 
where $\eta=\alpha-\beta$. In addition, the RG procedure yields a self-consistent equation
for the renormalized gap
\be
\Delta_r={\Delta_0\over [1+2\eta\ln E_c/\Delta_r]^{\alpha/\eta}}
\ee
the bare value of which is $\Delta_0(Q)=E^{(0)}_1(Q)-E^{(0)}_0(Q)=E_c(1-2Q/e)$.
 
As follows from (7), for $\eta>0$ the RG trajectory flows towards weak coupling
(see Fig.1) where the invariant charge $\tilde\Delta_r$ increases, although the actual gap
$\Delta_r$ given by (8) continues to decrease. In this regime, the 
quantum current fluctuations associated with the phnomenon of Coulomb blockade 
completely destroy the classical Josephson effect,
and the junction remains in the insulating state.
 This behavior appears to be similar to that emerging 
in the previously studied limit $\beta=0$ where
the system possesses the exact $XY$-symmetry and 
remains insulating for arbitrary values of the external charge $Q$ \cite{weak}.  

Near the line $\eta=0$ the RG flow slows down,
and right on that line the effective coupling $\alpha_r=\beta_r$ undergoes (almost) no
renormalization, while the effective splitting demonstrates a power-law
dependence on its bare value  
\be
\Delta_r={\Delta_0}({\Delta_0}/E_c)^{2\alpha/1-2\alpha},
\ee
which behavior is indicative of a possible 
dissipative phase transition at $\alpha=\beta=1/2$. 

Exactly at $\eta=0$ the system acquires the Ising symmetry, which suggests
that the transition in question is likely to be of the Kosterlitz-Thouless type \cite{weak}.
Moreover, since the effective coupling $\alpha=\beta$ undergoes no renormalization (at least, 
to first order), this critical behavior appears to be reminiscent of that 
of a junction with the $\it local$ Josephson energy and $\it Ohmic$ 
(as opposed to the quasiparticle tunneling-induced) dissipation.

In the regime $\alpha=\beta<1/2$, the quantum phase fluctuations are strong and
the energy bands $E_{0,1}(Q)$ remain non-degenerate, thereby forcing 
the junction to operate in the insulating (Coulomb blockade-dominated) regime. 
However, for $\alpha=\beta>1/2$  
the quantum fluctuations get quenched, and the energy bands 
become progressively more and more degenerate in a finite portion of the BZ which
expands from the boundaries $Q=\pm e/2$ inward as the parameter
$\alpha=\beta$ increases. 
This behavior indicates a suppression of the Coulomb blockade and a possible restoration
of the classical Josephson effect where the
voltage drop across the junction vanishes and the current is determined
by a (nonzero) average value of the phase difference $<\phi(\tau)>=const$ 
of the superconducting order parameters on the opposite banks of the junction.

In contrast, for $\eta<0$ one finds a runaway RG flow towards strong coupling (see Fig.1)
where both $\alpha_r$ and $\beta_r$ become of order unity
and $\tilde\Delta_r$ starts to decrease, regardless of the bare values of
these dissipative couplings. Such a behavior suggests that the junction 
is likely to remain in the (super)conducting regime for all $\alpha<\beta$.

Using the formula (8) for the gap, one can readily compute
the ground state energy $E_{0}(Q)=E_c-{1\over 2}\Delta_r(Q)$
which, in turn, yields the junction's effective capacitance
\be
C_r=({d^2E_0(Q)\over dQ^2})^{-1}|_{Q=0}\approx C[1+4\eta+O(\eta^2)]^{\alpha/\eta} 
\ee
Even more pronounced, appears to be 
the effect of tunneling on the average excess charge of the 
junction close to the BZ boundary ($\delta Q=Q-e/2\to 0^-$)
\be
<q(\tau)>=Q-C{dE_0(Q)\over dQ}={e\over 2}(1+{sign\delta Q\over
[1+2\eta\ln(e/\delta Q)]^{\alpha/\eta}}) 
\ee 
Consistent with the anticipated dissipation-induced suppression
of the Coulomb blockade, Eq.(11) demonstrates the rounding of the 
classic Coulomb staircase-like dependence of $<q(\tau)>$ on $Q$
with increasing $\alpha$.

Also, by continuing Eq.(8) analytically to the regime
$Q>e/2$ where the logarithmic function of $\delta Q$
acquires an imaginary part, one can estimate the decay rate of the first excited state
with energy $E_{1}(e/2-\delta Q)=E_0(e/2+\delta Q)$
\be
\Gamma(Q)=Im E_1(Q)\approx
{E_c}{2\pi\alpha(\delta Q/e)\over [1+2\eta\ln(e/\delta Q)]^{1+(\alpha/\eta)}}
\ee
The excited state remains well defined for as long as $\Gamma(Q)\ll\Delta_r(Q)$,
which condition can only be satisfied 
in the weak coupling regime ($\eta\ln(e/\delta Q)<1$).

Notably, the effects of the external charge screening and the reduction of the 
tunneling rate manifested by Eqs.(11) and (12), respectively, are both enhanced
as compared to the case of a normal junction ("single electron box") 
where $\beta=0$ \cite{weak}.

These two effects are also exhibited by the current-voltage ($I-V$) characteristics.
In a voltage-biased junction, the induced DC current 
\be
I(V)=<q(\tau)>\Gamma(Q)|_{Q=VC}\approx{2\pi\alpha(V-E_c/e)\over 
[1+2\eta\ln(E_c/eV-E_c)]^{1+2(\alpha/\eta)}}\theta(V-E_c/e)
\ee
vanishes below the threshold $V_c=E_c/e$. However, at finite temperatures
the hard Coulomb gap gets partially filled with thermally excited quasiparticle 
excitations, thus giving rise to a temperature-dependent conductance. By the same token, 
in a current-biased junction the zero-temperature $I-V$ characteristics are expected 
to become non-linear for $I < eE_c$.

Next, we consider the regime of strong dissipative couplings ($\eta>1$)
where the first insight into the problem can be obtained by virtue of an adaptation of
the variational technique. In this method, the correlation function
of the small ("spin wave-like") phase fluctuations is sought out in the form
\be
<|\phi_\omega|^2>={1\over \omega^2/E_c+g|\omega|+D}
\ee
The variational parameters $g$ and $D$ are then used to minimize the free energy 
$F=-T\ln {\cal Z}_0+T<S-\delta^2S_0>$ where $\delta^2S_0(\phi_\omega)$ is the quadratic action
which corresponds to Eq.(14) and determines the averages $<e^{i\phi}>=\exp(-{1\over 2}<\phi^2>)$.
In this way, one obtains a self-consistent equation for the effective dissipative coupling 
\be
\eta=g^{(g+1/g-1)}(2\beta)^{1/1-g}
\ee
It can be shown that if the following condition is fulfilled
\be
\alpha-\beta>e\theta(1/2-\beta)\ln{1\over 2\beta}
\ee
[here $e$ is the Euler's number], there exists a finite gap due to the $XY$-anisotropy in the two-dimensional
vector space spanned by the unit vector ${\bf n}=(\cos \phi/2,\sin \phi/2)$
\be
D=E_c(2\beta)^{g/g-1}g^{2/1-g}
\ee
where $g\approx\eta^{1-2/\eta}(2\beta)^{1/\eta}$ is the solution of Eq.(15) for $\eta>0$.
For comparison, within almost the entire domain $\eta<0$ one finds $g\approx \eta$. 
  
The gap (17) attains its bare value $2\beta E_c$ at $g\to\infty$ and
decreases upon lowering the energy cutoff. Considering that 
the bandwidth of the phase fluctuations, too, 
undergoes a downward renormalization from its bare value $E_c$ (see Eq.(28) below),
one finds that the spectrum of the quadratic operator $\delta^2S_0/\delta\phi^2$ 
remains stable for both positive and negative $\eta$.

It should also be noted that, when continued to real frequencies,
the spectrum of the small phase fluctuations appears to be strongly 
overdamped. Therefore, except for the immediate vicinity of the separatrix $\eta=0$
(hence, in almost the entire $\alpha-\beta$ plane),
it bears no resemblance
to the simple plasmon pole (cf. Ref.\cite{joglekar}).

In the presence of a nonzero $XY$-anisotropy $D\neq 0$, the expectation 
value $<\phi^2(\tau)>$ becomes finite and the phase $\phi$ gets localized. As a result, 
the real part of the AC conductance
\be
G(\omega)=I(\omega)/V(\omega)=Re{1\over \omega_n
<|\phi_{\omega_n}|^2>}|_{\omega_n\to -i\omega}
\ee
develops a coherent peak at zero frequency, $G(\omega\to 0)=D\delta(\omega)$,
thus indicating the onset of the classical Josephson effect. 

In the case that Eq.(15) features no solution, 
the gap $D$ vanishes and the expectation value of the phase fluctuations
$<\phi^2(\tau)>$ diverges, so that the phase remains delocalized. 
At the first sight, it may seem that the real part of the conductance 
approaches a finite value $G(0)=g$ in the DC limit. However, we expect that
the continuing downward renormalization of the effective dissipative parameter $\eta$ 
(see Eq.(19) below) will eventually bring the system into the insulating regime.

In order to justify the above claim, we point out that is in the absence of a non-trivial mean field solution
the variational method must be abandoned 
in favor of a straightforward perturbative approach.
The latter yields a continuous renormalization
of the dissipative couplings due to the non-Gaussian (quartic and higher order) terms
in the action (1). 

To the leading order, such a renormalization is described by the RG equations
\be
{d\alpha_r\over d\ln\tau_c}=-{\alpha_r\over  2\pi^2\eta_r},~~~~~
{d\beta\over d\ln\tau_c}=-{\beta_r\over  2\pi^2\eta_r}
\ee
which demonstrate a monotonic decrease of the effective dissipative
parameter as a function of the lowered energy cutoff $1/\tau_c^\prime$
\be
\eta_r=\eta-{1\over 2\pi^2}\ln {E_c\tau_c^\prime}
\ee 
It is worth noting that the steady decrease of both $\alpha_r$ and $\beta_r$ 
as well as the approximate constancy of the ratio $\alpha_r/\beta_r$ manifested by Eqs.(19) 
are consistent with the behavior found in the weak coupling regime, and so the 
RG trajectories interpolate smoothly between the weak and strong coupling regimes.

Interestingly enough, the RG equations 
(19) coincide with those of the spin-boson model
\be
S_{sb}[{\bf m}(\tau)]={i\over 2}\int d\tau(1-m_z)
{\partial\phi\over \partial\tau} +\int d\tau({1\over 4}E_cm_z^2+m_xh_x+m_yh_y)
\ee
where ${\bf m}=({\sqrt {1-m_z^2}}\cos\phi/2,{\sqrt {1-m_z^2}}\sin\phi/2,m_z)$ 
is a 3D unit vector field composed of the phase $\phi$
and a conjugate "momentum" $m_z$. The first term in (21) is
the spin's Berry phase, and  a two-component
random field $h_{x,y}(\tau)$ represents a dissipative bath governed by the correlators 
$<|h_x(\omega)|^2>=(\alpha+\beta)|\omega|$ and $<|h_y(\omega)|^2>=\eta|\omega|$. 

The origin of the equivalence between the strong coupling regimes of the actions (1,3)
and (21) can be traced back to the observation that at large $\eta$ 
both, the discreteness of charge
and the quantum nature of spin (or, for that matter, the spin's Berry phase),
turn out to be largely irrelevant and the actions (1,3) and (21)
are both dominated by the (identical) dissipative terms. 

As another side note, we mention that the action (21) 
can also be encountered in generic (spin anisotropic)
models of quantum impurities as well as those of noisy 
qubits (generalized (pseudo)spin-$1/2$ systems exposed to dissipative environments).
In light of the earlier observations \cite{vojta} that in
such models the only stable fixed points tend to be those of the Ising symmetry, 
it does not seem unreasonable to surmise that the behavior in the entire region defined by
Eq.(16) can, in fact, be similar to that found for $\alpha=\beta>1/2$.

Altogether, the above results suggest a tentative layout of the 
phase diagram where the insulating behavior 
sets in throughout the domain defined by the 
inequalities $0\leq\beta<1/2$ and $\beta\leq \alpha\leq\beta+e\ln(1/2\beta)$ (see Fig.1).
By implication, we then conjecture
that upon increasing the value of $\eta>0$ or changing its sign, the insulator gives way
to the (super)conducting phase. 

In order to further refine our understanding of the strong coupling 
regime we employ the semiclassical approach in the phase representation \cite{strong}. 
The Coulomb energy (which, contrary to the weak coupling case, is now subdominant
to the dissipative terms) induces a splitting between 
the (otherwise degenerate) phase configurations ("vacua"), and the partition function 
\be
{\cal Z}(Q)=\sum_{N=0}^\infty
\int_0^{2\pi}d\phi_0\int^{\phi_0+4\pi N}_{\phi_0}D\phi(\tau)e^{-S[\phi(\tau)]+2\pi iNQ/e}
\ee
is now saturated by the trains of $4\pi$-phase slips ("instantons") 
corresponding to the transitions between 
different even/odd vacua $\phi_{2M}=2\pi M$ and $\phi_{2M+1}=(2M+1)\pi$.

A viable candidate to the role of the trajectory connecting the vacua
$\phi_{2M(2M+1)}$ and $\phi_{2M+N(2M+1+N)}$ can be chosen in the form
(here $q_k=\pm 1$ and $\sum_{k=1}^nq_k=N$)
\be
\Delta\phi_{N}(\tau)=4\sum^n_{k=1}q_k\tan^{-1}\Omega_k(\tau-\tau_k)
\ee
which is inspired by the exact solution of the classical equations of motion for
$\beta=0$ and $E_c=\infty$.  
 
Evaluating the action (1,3) on the single instanton ("bounce") trajectory  
$\Delta\phi_{1}(\tau)$ we obtain
\be
S_1(\Omega)=2\pi^2\eta+{\pi\Omega\over 2E_c}+{4\pi\beta E_c\over \Omega}
\ee
Minimizing Eq.(24) with respect to the instanton's size $\Omega$, we find
$\Omega_{min}\sim E_c\beta^{1/2}$ which is small compared to the average spacing 
between the instantons ($|\tau_n-\tau_m|\sim 1/NT$) for $T\to 0$.
                        
Since the instantons are well separated, one can expand about a single-instanton configuration 
($\phi(\tau)=\Delta\phi_{1}(\tau)+\delta\phi(\tau)$) and obtain
a quadratic action of the Gaussian fluctuations 
\bea
\delta^2S_1=T\sum_\omega
(\alpha|\delta\phi_\omega|^2[|\omega+\Omega/2|+|\omega-\Omega/2|-|\Omega|]+
\beta[\delta\phi_\omega\delta\phi^*_{\Omega-\omega}
(E_c/T-|\omega+\Omega/2|-\Omega/2)\nonumber\\
+\delta\phi_\omega\delta\phi^*_{\Omega+\omega}
(E_c/T-|\omega-\Omega/2|-\Omega/2)])
\eea
At $\beta=0$ the eigenvalues of the operator
$\delta^2S_1$ with the frequencies $|\omega|<\Omega$ turn out
to be independent of $\alpha$. However, 
for any finite $\beta$ these "soft" modes acquire gaps 
and, therefore, can not be separated from the rest of the spectrum (cf. \cite{strong}). 
The fact that this spectrum remains 
positively defined and, therefore, stable justifies the use of Eq.(24)
as a sensible approximation for the (unknown) true local minimum of the action (1,3). 

Next, by singling out the zero modes 
corresponding to the changes in the instanton position
($\delta\phi^{(1)}_1\propto\partial\phi_1/\partial\tau$) and size
($\delta\phi^{(2)}_1\propto\partial\phi_1/\partial\Omega$)
and integrating over the non-zero normal modes, one can cast
the partition function as that of a dilute instanton gas 
\be
{\cal Z}(Q)\approx\sum_{N=1}^\infty{1\over N!}
\prod_{k=1}^N\int^{1/T}_0d\tau_k\int {d\Omega_k}J(\Omega_k)
{\big (}{\det\delta^2S_0/\delta\phi^2\over \det\delta^2S_1/\delta\phi^2}{\big )}^{1/2}
e^{-S_1(\Omega_k)+2\pi iNQ/e}
\ee
where $J(\Omega)$ is the measure of integration over the zero modes,
and the determinant stemming from the integration over the non-zero modes
is normalized with respect to that of the operator $\delta^2S_0/\delta\phi^2$ 
corresponding to the Gaussian fluctuations about the trivial vacuum $\phi(\tau)=0$.

At large positive values of $\eta$ the ground state energy can be obtained by taking the
zero temperature limit of the logarithm of Eq.(26)
\be
E_0(Q)=-T\ln {\cal Z}(Q)|_{T\to 0}=E_c\cos(2\pi Q/e)
\int d\Omega J(\Omega)
{\big (}{\det\delta^2S_0/\delta\phi^2\over \det\delta^2S_1/\delta\phi^2}{\big )}^{1/2}
e^{-S_1(\Omega)}
\ee
At $Q=0$ Eq.(27) yields a renormalized bandwidth of the phase fluctuations 
whose asymptotic behavior for $\eta\gg 1$ reads 
\be
E_{c,r}\sim E_c\eta e^{-2\pi^2\eta}
\ee
By analogy with the $\beta=0$ case, this reduction of the bandwidth
can be attributed solely to the exponential blow-up of the effective capacitance
$C_r\sim C\exp(2\pi^2\eta)/\eta$. For comparison, 
along the separatrix $\alpha=\beta$, the effective 
capacitance grows only as $C_{r}\sim C\exp(\pi{\sqrt \alpha}/8)/\alpha^{1/4}$
for large $\alpha$.

The advantage of working in the phase representation is that
one can better elucidate the nature of the putative dissipative phase transition
by relating it to a spontaneous breaking of the symmetry between 
the $\phi_{even}$ and $\phi_{odd}$ vacua \cite{strong}.

In the disordered phase, the correlation 
function $<e^{i\phi(\tau)/2}e^{-i\phi(0)/2}>$ decays algebraically
and the insulator-like $I-V$ characteristics show the 
presence of a hard (at $T=0$ and $I\to 0$) Coulomb gap.  
By contrast, in the ordered phase the system develops an 
order parameter $<e^{i\phi(\tau)/2}>\neq 0$ whose presence indicates 
that the phase variable is localized in either even or odd vacua.
This does not, however, constitute a complete phase localization, and, therefore, 
the apparent classic Josephson effect should be considered
a non-equilibrium phenomenon which can only be observed 
at finite (albeit potentially quite long) observation times, while at still longer times
the response would revert to the resistive behavior \cite{sz}.

Another potentially important non-equilibrium effect is excitonic enhancement of the
tunneling probability which modifies the exponent in the power-law in Eqs.(3) to 
$2-\epsilon$ where $\epsilon$ is a function of the (non-universal) 
scattering phase shift \cite{arovas}.
In this "sub-Ohmic" case, the right-hand-sides of the RG equations (19) pick up additional terms $\epsilon\alpha$ and $\epsilon\beta$.
Consequently, one now finds a fixed point at $\eta_c=1/2\pi^2\epsilon$, 
upon approaching which the bandwidth vanishes as 
$E_c^*\sim E_c(1-\eta/\eta_c)^{(1-\epsilon)/\epsilon}$.
One would then be led to the conclusion that in the domain defined as
$\eta_c\leq\alpha-\beta\leq e\ln(1/2\beta)$ the transition from the insulating to  
the Josephson phase is preempted by that into a new conducting 
(albeit, possibly, dissipative) state. 

We anticipate that the prediction of the dissipative phase transition
made in this work might be possible
to test by using junctions between thin single-crystal cuprate films.
For the parameter values found in Ref.\cite{joglekar}
($\alpha=2\beta$ and $E_c\sim\Delta_d$)
one would expect to observe the apparent superconducting response 
at $\alpha>\alpha_c\approx 0.85$. 
Considering that the typical junctions manufactured from the cuprate 
superconductors tend to have relatively high conductances
\cite{hm}, this condition can be readily met.

A still more realistic model of the cuprate films
would also feature a renormalized kernel $\alpha(\tau)$ that accounts for 
possible zero-energy bound states (which is only 
absent in the case of the node-to-node tunneling) 
as well as a quadratic Ohmic dissipative term representing the effect 
of a shunting normal resistance (metallic environment 
surrounding the superconducting grains). We anticipate that
by including these effects into the model (1,3) 
the borderline between the (super)conducting and insulating
phases can be pushed even further towards lower values of the dissipative couplings.

In contrast to our findings, 
the naive gradient expansion employed in Ref.\cite{joglekar} is not 
capable of revealing any critical coupling $\alpha_c$
or the corresponding phase transition in question. 
In fact, a direct application of this technique
to the model (1,2) would yield 
an undamped correlation function $<|\phi_\omega|^2>=(\omega^2/E^*_c+D^*)^{-1}$ with 
a strongly temperature-dependent gap 
$\Delta\sim (E_cT)^{1/2}$ throughout the entire $\alpha-\beta$ plane
\cite{joglekar}, which behavior is starkly different from that found in this work. 

Also, as far as a proper modeling of the 
highly inhomogeneous cuprate samples studied in Refs.\cite{hm}
is concerned, the latter are likely to be best 
described not by the strictly two-dimensional action (1,3)
but rather as a network of three-dimensional grains
whose behavior is expected to be somewhat more conventional \cite{sz}. 

To conclude, in this work we carried out a comprehensive analysis of the 
quantum dynamics of a node-to-node Josephson junction between two planar 
d-wave superconductors. Our results reveal the presence of a dissipative phase 
transition in the $\alpha-\beta$ plane.
We also discussed properties of the physical observables
that are indicative of the transition in question,
including the junction's 
energy spectrum, effective capacitance and $I-V$ characteristics.

The corresponding critical behavior differs, in a number 
of important aspects, from the previously studied cases of a superconducting
junction with the local Josephson energy as well as that of a normal juncion. 
However, in a general agreement with the earlier analyses of these systems
the presence of a sufficiently strong quasiparticle $and$ Cooper pair tunneling 
turns out to be instrumental for observing the Josephson behavior (albeit, possibly, 
not at the longest time scales). 

This research was supported by NSF under Grant DMR-0349881 and ARO under 
Contract DAAD19-02-1-0049.

\begin{figure*} 
	\includegraphics[height=1.0\textwidth,angle=0]{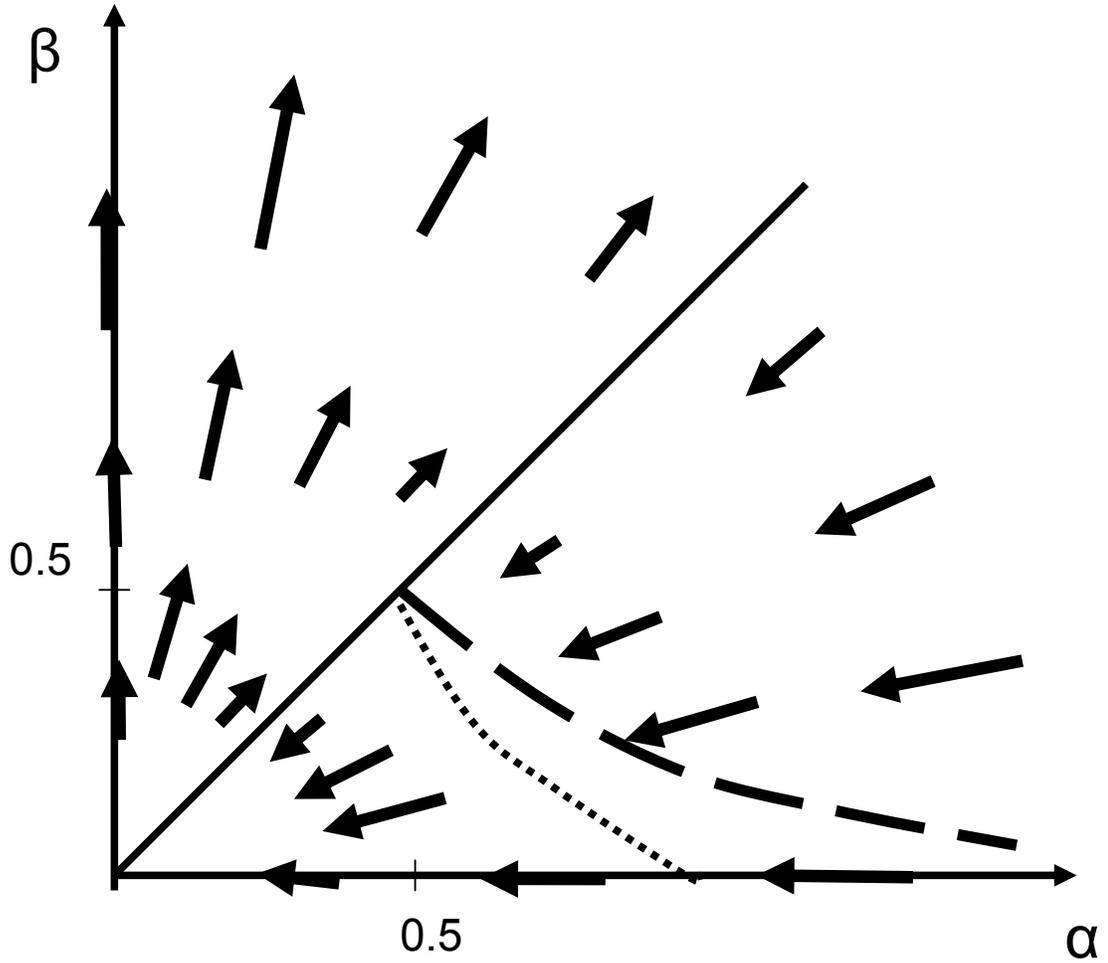} \vspace{0.3cm}
	\caption {\label{Fig1} Phase diagram of the anisotropic $XY$-model.
Renormalization group trajectories are shown by arrows,
and the dashed (dotted) line represents the phase boundary 
between insulating and (super)conducting phases in the case of Ohmic (sub-Ohmic) dissipation.}
\end{figure*}

\end{document}